\documentclass[pra,amsmath,amssymb,twocolumn,superscriptaddress,hidepacs]{revtex4}
\usepackage{CJK}
\usepackage{amsmath,amssymb}
\usepackage[usenames]{color}
\usepackage{amssymb}
\usepackage{grffile}
\usepackage[pdftex]{graphicx}
\usepackage{amsmath, amstext, amssymb, amsfonts, amsxtra}
\usepackage{textcomp}
\usepackage{xspace}
\usepackage{epstopdf}
\DeclareGraphicsExtensions{.eps}
\usepackage[percent]{overpic}
\usepackage{enumerate}
\usepackage{mathbbol}
\usepackage{graphicx}
\usepackage{bm,amsmath,amssymb}
\usepackage{dcolumn}
\usepackage{mathrsfs}
\usepackage{cases}
\usepackage{color}
 \usepackage{ulem}
 \usepackage{indentfirst}
 \usepackage{diagbox}

\newcommand{\be}{\begin{equation}}
\newcommand{\ee}{\end{equation}}
\newcommand{\bea}{\begin{eqnarray}}
\newcommand{\eea}{\end{eqnarray}}

\begin{document}
\begin{CJK*}{GBK}{song}

\title{Off-site trimer superfluid on a one-dimensional optical lattice\thanks{Project supported by the National Natural Science Foundation of China (Grant Nos.~11305113), and the project GDW201400042 for the ``high end foreign experts project''.}}
\author{Fan Er-Nv}
\affiliation{College of Physics and Optoelectronics, Taiyuan University of Technology Shanxi 030024, China
}
\author{Scott C Tony}
\affiliation{College of Physics and Optoelectronics, Taiyuan University of Technology Shanxi 030024, China
}
\affiliation{
Near India Pvt Ltd, no. 71/72, Jyoti Nivas College Road, Koramangala, Bengalore 560095, India
}
\author{Zhang Wan-Zhou}
\thanks{Corresponding author. E-mail:~zhangwanzhou@tyut.cn}
\affiliation{College of Physics and Optoelectronics, Taiyuan University of Technology Shanxi 030024, China
}

\date{\today}
\begin{abstract}
The Bose-Hubbard model with an effective off-site three-body tunneling,
characterized by jumps towards one another,
between one atom on a site and a pair atoms on the neighborhood site,  is studied systematically on a one-dimensional lattice, by using the density matrix renormalization group method. The off-site trimer superfluid, condensing  at momentum $k=0$, emerges in the softcore Bose-Hubbard model but it disappears in the hardcore Bose-Hubbard model. Our results numerically verify that the off-site trimer superfluid phase derived in the momentum space from [Phys. Rev. A {\bf 81}, 011601(R) (2010)] is stable in the thermodynamic limit.
The off-site trimer superfluid phase, the partially off-site trimer superfluid phase and the Mott insulator phase are found, as well as interesting phase transitions, such as the continuous or first-order phase transition from the trimer superfluid phase to the Mott insulator phase.
Our results are helpful in realizing this novel off-site trimer superfluid phase by cold atom experiments.
\end{abstract}
%\textbf{Keywords:} Bose-Hubbard model, off-site trimer superfluid, density matrix renormalization group method
%\textbf{PACS:} 37.10.Jk, 05.30.Jp, 03.75.Lm
\maketitle

\section{introduction}
Exploring the novel phases of cold atom systems is an important guide in understanding the collective behavior of the quantum many-body system.
The optical lattice \cite{opt1, opt2, opt3} provides us an ideal and perfect platform hosting the cold atoms,
described by the Bose-Hubbard (BH) model.\cite{djak, mgre}
Interestingly, various kinds of novel quantum phases, such as the atom superfluid (ASF) phase,\cite{sf1, sf2}
the pair SF (PSF) phase or the trimer SF (TSF) phase, emerge on the optical lattices, by controlling laser as well as other various parameters.
Also there are interesting mixed phases, such as the partially paired superfluid phase.\cite{pp,ppphase}

In the ASF phase, one atom jumps repeatedly to its neighboring sites. In the PSF phase, a pair of atoms are able to tunnel simultaneously from one site to the neighborhood sites, driven by the attractive interaction,\cite{cqst5, cqst6, thermal} or laser-assisted to the state dependent optical latice.\cite{xfzhou}
Another type of PSF phase is characterized by a jump of the off-site pair atoms, from the sites $i$ and $j$ to another two sites labeled $k$ and $l$.\cite{hcjiang, corre} In spin language, the off-site pair tunneling can be represented in the spin-$\frac{1}{2}$ XY model with four-spin interactions.\cite{s12, fourspin, fspin}

Naturally, the question arises: can the off-site trimer superfluid(OTSF) exist in a stable fashion?
In the OTSF phase, a trimer is composed of one atom on a site and the other two atoms on the neighborhood site.

Fortunately, the derivation of the Hamiltonian with an effective off-site trimer model in Ref.~\cite{central}, shows that the off-site trimer model exists probably in theory.
Moreover, condensation of the off-site trimer on a kagom\'{e} lattice\cite{zhaihui} supports the existence of the OTSF phase in quantum many-body systems.
Furthermore, experimental realization of local trimers, by using 7Li,\cite{li7, li7b, li7c} 39K,\cite{39K} 85Rb\cite{85Rb} and 133Cs,\cite{133Cs, 133Csb} and numerical studies of trimer-tunneling model\cite{liran, yangyuan} are helpful in understanding the OTSF phase.

Herein, we use the density matrix renormalization group (DMRG) method\cite{dmrg, dmrg2} to study a BH model with off-site trimer tunneling on a one-dimensional optical lattice, and find that the OTSF phase derived in the momentum space from Ref.~\cite{central} exists in a stable fashion in the thermodynamic limit.
This OTSF phase emerges in the softcore BH model though it disappears in the hardcore model. Interestingly, the partial OTSF (POTSF) phase, i.e. a phase mixed with the ASF and OTSF phases, is also found.

The outline of this work is as follows.
Section~\ref{sec:model}~~ shows the Hamiltonian model.
Section~\ref{sec:method}~~ describes the DMRG method and the useful observables.
Section~\ref{sec:result}~~~ studies both the hardcore and softcore BH models, and then finds out if the OTSF  and POTSF phases exist in the softcore BH model, and the phase diagrams are also discussed.
Concluding comments are made in Sec~\ref{sec:con}.
\section{The Hamiltonian model}
\label{sec:model}
\begin{figure}[hbt]
\begin{center}
\includegraphics[width=0.8\columnwidth]{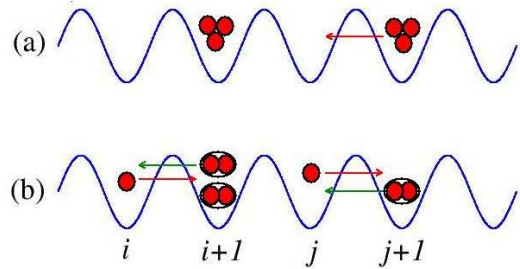}
\caption{(color online) (a) A static trimer on the site labeled $i+1$ and an active trimer tunneling between $j$ and $j+1$.  (b) An off-site trimer  tunneling between $j$ and $j+1$ in the hardcore model  with $n_{max}^{p}=1$ and an
off-site  trimer with a softcore model with $n_{max}^{p}=2$ between $i$ and $i+1$.}
\label{off-fig}
\end{center}
\end{figure}

The starting point is the BH Hamiltonian with an off-site trimer tunneling
term, as shown in Ref.~\cite{central}
\be
H=H_{a}+H_{p}+H_{ap},
\label{H}
\ee
with
\be
\begin{aligned}
H_a=& -J\sum_{i=1}^L\left(a_i^\dagger a_{i+1}+H.c.\right)+\sum_{i=1}^Lh_i^a,
\end{aligned}
\ee
where
$J$ is the atom hopping rate,  $a_i^\dagger$ ($a_i$)
is the atom creation (annihilation) operator, the operator on each site is $h_i^a=\frac{U}{2}(n_i^a(n_i^a-1))-\mu_a n_i^a$ and $n_i^a=a_i^\dagger a_i$ is the occupation number operator at site $i$,
$U$ is the on-site two-body interaction energy and $\mu_a$ is the chemical potential of  atoms (for simplicity, we ignore the subscript ``$a$'' in this work).
$H_a$ describes the single atom tunneling and the interactions between atoms.

The second term $H_p$ exhibits the tunneling and interactions
between bounded pairs of atoms and is given by
\be
\begin{aligned}
H_p=&-J_p\sum_{i=1}^L\left(p_i^\dagger p_{i+1}+H.c.\right)+\sum_{i=1}^Lh_i^p,
\end{aligned}
\ee
with $p_i^\dagger$ ($p_i$) being the pair (bound-particle pair) creation (annihilation)
operator, $h_i^p=\frac{U}{2}(n_i^p(n_i^p-1))-\mu_pn_i^p$ and $n_i^p=p_i^\dagger p_i$ is the number operator of paired atoms at site $i$.
The effective pair hopping is $J_p=-\frac{2J^2}{U}$,\cite{central}
and $\mu_p=U-2J_{p}$.

$H_{ap}$ describes the effective short-range interactions between the single atom and the paired atoms, as follows,
\be
H_{ap}=V\sum_{i=1}^L n_i^p n_{i+1}^a-W\sum_{i=1}^L\left(p_{i}a_i^\dagger p_{i+1}^\dagger a_{i+1}+H.c.\right),
\label{Hap}
\ee
where $V=-\frac{7J^2}{2U}$ denotes a weak attractive nearest-neighbor interaction, and $W=2J$\cite{central} is the off-site trimer tunneling energy, as shown in Fig.~\ref{off-fig}(b).

It should be noted that particles ``a'' and ``p'' are the same kind of atoms.
Here, ``a'' denotes the single atom component and ``p'' represents the component of the paired atoms.

\section{Methods and order parameters}
\label{sec:method}
\begin{figure}[hbt]
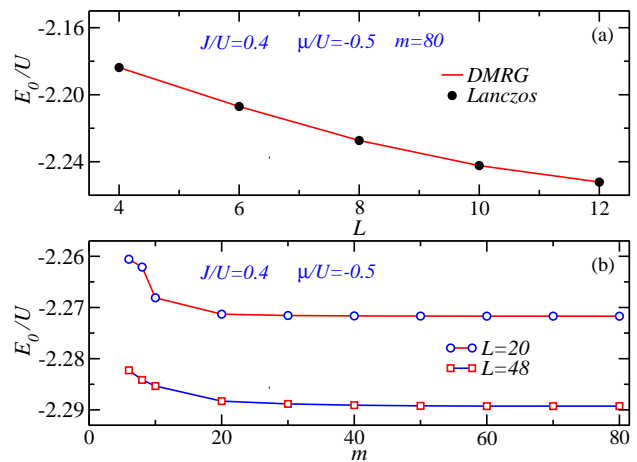

\begin{center}
\includegraphics[width=0.95\columnwidth]{fig2a.eps}\\
\includegraphics[width=0.95\columnwidth]{fig2b.eps}
\caption{(color online) The ground state energies of the hardcore BH model with $n_{max}^{a}=n_{max}^{p}=1$ at $J/U=0.4$, $\mu/U=-0.5$ (a) by the Lanczos and DMRG methods with different sizes $L$. And in the DMRG method, the dimension of the truncated matrix takes $m=80\sim130$. (b) The ground state energies versus $m$ with sizes $L=20$ and $48$.}
\label{LD}
\end{center}
\end{figure}

To explore the OTSF phase and get the reliable phase diagrams, we use the DMRG method \cite{dmrg, dmrg2, dmrg3} to simulate the model described by Eq.~(\ref{H}) for a chain with a periodic boundary condition. In order to ensure the correctness of the DMRG method, we compare the ground state energies of the hardcore BH model with $n_{max}^{a}=n_{max}^{p}=1$ by the Lanczos and DMRG methods.
The chosen parameters are $J/U=0.4$, $\mu/U=-0.5$, and the dimension of the truncated matrix takes $m=80\sim130$ within the DMRG method.
The results from both methods are completely the same within $10$ digits with $L=4, 6, 8, 10$, and  the first $5$ significant numbers are given in Tab. \ref{Tab:t1}. While $L=12$, data from both methods are the same within $4$ digits. The consistency can be seen in Fig.~\ref{LD}(a).

The dimensions $m=80\sim130$ of the truncated matrix are reliable to get the ground state of the Hamiltonian. The ground state energies versus $m$ with sizes $L=20$ and $48$ are shown in Fig.~\ref{LD}(b) and they are convergent when $m=80$. Furthermore, in the present work, the attractive interaction considers $V\textless0$, and the number of possible phases is smaller than that with repulsive interaction, where various density wave (DW) phases emerge in the extended BH model on a one-dimensional optical lattice.\cite{dw1, dw2, dw3} Therefore, $m=80\sim130$ is reasonable for the parameters with $V\textless0$ and the results seem easier to converge.

\begin{table}[hbt]
\caption{Values of the ground state energies by the Lanczos and DMRG methods with different sizes $L$.}
\begin{center}
\begin{tabular}{c|cccccc}
  \hline
  \hline
  \diagbox[dir=SE]{Methods}{L} &4&6&8&10&12\\   \hline
 Lanczos &-2.18378&-2.20702&-2.22733&-2.24232&-2.25214\\
 DMRG    &-2.18378&-2.20702&-2.22733&-2.24232&-2.25212\\
 \hline
 \hline
\end{tabular}
\label{Tab:t1}
\end{center}
\end{table}

The average correlation functions\cite{hcjiang} are defined to distinguish each quantum phase as follows,
\begin{equation}
\begin{aligned}
C_{a}&=\frac{\sum_{r=1}^{L}C_{a}(r)}{L}, C_{a}(r)=\langle a_i^\dagger a_{i+r}\rangle, \\
C_{p}&=\frac{\sum_{r=1}^{L}C_{p}(r)}{L}, C_{p}(r)=\langle p_i^\dagger p_{i+r}\rangle, \\
C_{ap}&=\frac{\sum_{r=1}^{L}C_{ap}(r)}{L}, C_{ap}(r)=\langle p_{i}^\dagger a_{i}a_{i+r}^\dagger p_{i+r}\rangle.
\end{aligned}
\end{equation}

With finite sizes, the ASF phase is characterized by $C_{a} \ne 0$ and
the OTSF phase is denoted by  $C_{ap}$. For the POTSF phase, both $C_{a}$  and  $C_{ap}$
are nonzero.
Although the average correlation functions $C_a$, and $C_{ap}$ tend to be
zero in the thermodynamic limit, i.e., $L\to\infty$, $\int_{0}^{L}C_a(r)/L=0$ and $\int_{0}^{L}C_{ap}(r)/L=0$,
we can still use  $C_a$, and $C_{ap}$ with the finite sizes to conveniently judge  phase transitions.
Finally, by $C_a(r)$ and $C_{ap}(r)$
described by algebraically or exponentially decaying functions, we can determine whether or not
the novel phases really exist in the thermodynamic limit. The on-site interaction $U$ is taken as a unit $U=1$, and the chemical potential of pairs, i.e. $\mu_p=U-2J_p=U-2\times(-\frac{2J^2}{U})\textgreater1$ is relatively larger. Hence, for both the hardcore or softcore models,  the system sits in the MI phase, i.e., MI($\rho_p=1$) for the hardcore model and MI($\rho_p=2$) for the softcore model, where  the quantities $C_{p}(r)$ and  $C_{p}$ remain at zero, and therefore the pair momentum distribution $n_p(k)$ has no peak.
Herein, we ignore the detailed discussion about the  corresponding quantities ( $C_p(r)$, $C_p$ and $n_p(k)$).

Furthermore, the momentum distributions are also used to detect various SF phases and defined as \cite{momentum}
\begin{equation}
\begin{aligned}
n_{a}(k)&=\frac{1}{L}\sum_{i,j}\langle a_i^{\dagger}a_{j}\rangle e^{ik(i-j)}, \\
n_{p}(k)&=\frac{1}{L}\sum_{i,j}\langle p_i^{\dagger}p_{j}\rangle e^{ik(i-j)}, \\
n_{ap}(k)&=\frac{1}{L}\sum_{i,j}\langle p_{i}^\dagger a_{i}a_{j}^\dagger p_{j}\rangle  e^{ik(i-j)}.
\end{aligned}
\label{momentum}
\end{equation}

The momentum distributions will have peaks for the three types SF phases. Specifically, $n_{a}(k)$ will peak at $k=0$ in the POTSF and ASF phases, and $n_{ap}(k)$ will peak at $k=0$ in the POTSF and OTSF phases.

Moreover, the integer fillings of the single atom component $\rho_a=\frac{\sum_{i=1}^{L}\langle n_i^a\rangle}{L}$
and the paired-atom component $\rho_p=\frac{\sum_{i=1}^{L}\langle n_i^p\rangle}{L}$  are also used to characterize the Mott insulator (MI) phase.
In Tab. \ref{Tab:t2}, we list the values of the order parameters for several typical phases, namely the OTSF, ASF and MI phases.
 The ASF phase means that the system sits in  the  SF$_a$+MI$_p$  phase.

\begin{table}[hbt]
\caption{Values of the order parameters $C_{a}$ and $C_{ap}$, and characteristics of $C_a(r)$, $C_{ap}(r)$, $n_a(k)$ and $n_{ap}(k)$ for typical phases. ``$A$'' and ``$E$'' denote $C_{a}(r)$ or $C_{ap}(r)$ obeys
a decaying behavior, algebraically and exponentially, respectively. ``$\surd$'' means $n_{a}(k)$ or $n_{ap}(k)$ will peak at $k=0$. }
\begin{center}
\begin{tabular}{c|ccccccc}
  \hline
  \hline    ~~&~~  ~~$C_a$~~  ~~&~~ ~~$C_{ap}$~~ ~~&~~  ~~$C_a(r)$~~ ~~&~~ ~~$C_{ap}(r)$~~ ~~&~~ ~~$n_a(k)$~~ ~~&~~ ~~$n_{ap}(k)$~~\\ \hline
  POTSF          &   $\ne0$            & $\ne0$    &   $A$            & $A$   &   $\surd$            & $\surd$  \\
  OTSF           &   0                 & $\ne0$    &   $E$            & $A$   &   $     $            & $\surd$\\
  ASF            & $\ne0$              &   0       &   $A$            & $E$   &   $\surd$            & $     $ \\
  MI             & 0                   &   0       &   $E$            & $E$   &   $     $            & $     $ \\
  \hline
  \hline
\end{tabular}
\label{Tab:t2}
\end{center}
\end{table}

\section{Results}
\label{sec:result}
\subsection{Hardcore off-site trimer BH model}
\begin{figure}[hbt]\vskip 0.5cm
\begin{center}
\includegraphics[width=0.9\columnwidth]{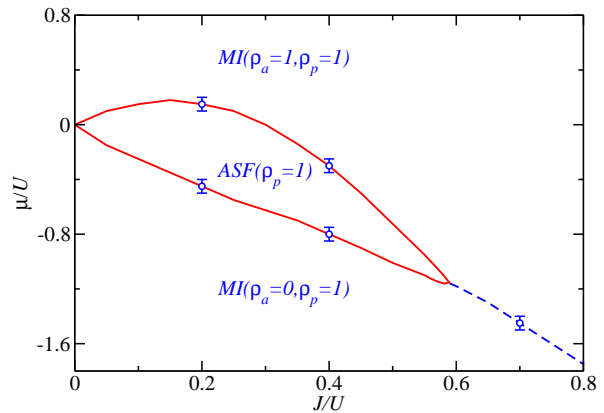}
\caption{(color online) The hardcore phase diagram with $n_{max}^{a}=n_{max}^{p}=1$, which contains the MI($\rho_a=0$, $\rho_p=1$), ASF and MI($\rho_a=1$, $\rho_p=1$) phases in the plane ($J/U$, $\mu/U$). The lines are schematic phase boundaries. The dashed and solid lines denote the first order and continuous phase transitions, respectively.}
\label{n1md}
\end{center}
\end{figure}
Initially, we studied the hardcore BH model with the maximum occupation number $n_{max}^{a}=n_{max}^{p}=1$, where each site is only allowed to be occupied by  one atom or a pair of atoms.

The phase diagram is  shown in Fig.~\ref{n1md}, where no OTSF phase exists and only the ASF and MI phases emerge.
The system sits in the MI($\rho_a=0$, $\rho_p=1$) phase if $\mu/U\textless0$, and the system enters into the MI($\rho_a=1$, $\rho_p=1$) phase when $\mu/U$ become larger. The ASF phase, i.e., the SF$_a$+MI$_p$ phase exists in the middle of the phase diagram,
and is similar to the SF$_{A}$+MI$_{B}$ phase in Ref\cite{sfmi}. The phase transitions from the ASF phase to the MI phases are continuous as expected.

\begin{figure}[t]
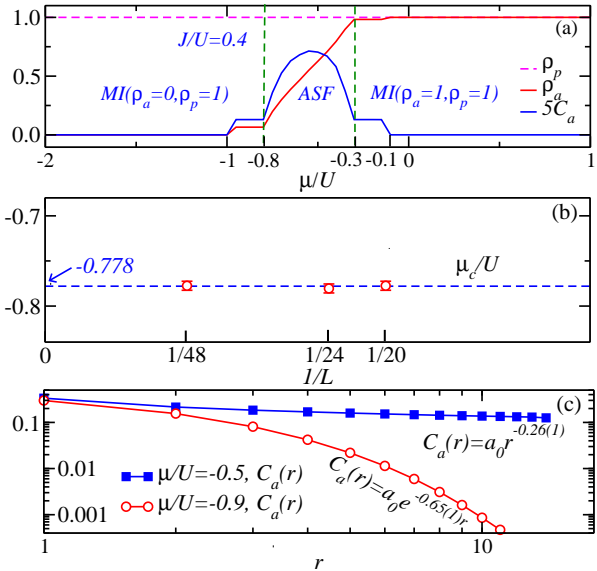

\begin{center}
\includegraphics[width=0.9\columnwidth]{fig4a.eps}\\ \hskip -0.15cm
\includegraphics[width=0.905\columnwidth]{fig4b.eps}\\
\includegraphics[width=0.9\columnwidth]{fig4c.eps}
\caption{(color online) (a) The values of the densities $\rho_a$, $\rho_p$ and average correlation function $C_a$ along $\mu/U$ at $J/U=0.4$ with size $L=48$. (b) The phase transition point $\mu_c/U$ vs $1/L$ from the MI($\rho_a=0$, $\rho_p=1$) phase to the ASF phase.
(c) Correlation $C_{a}(r)$ at $\mu/U=-0.9$ and $-0.5$.}
\label{N1J04}
\end{center}
\end{figure}
For a more detailed description of the each physical quantity, we choose to  scan  $\mu/U$ at $J/U=0.4$ as shown in Fig.~\ref{N1J04}.
In Fig.~\ref{N1J04}(a), when $\mu/U\textless-1$, the single-atom component sits in the empty phase, i.e., $\rho_a=0$.
The paired-atom component sits in the MI phase, i.e., $\rho_p=1$. Due to the relationship,
\be
\mu_p=U-2J_p=U-2\times(-\frac{2J^2}{U})\textgreater1,
\label{mup}
\ee
$\mu_p$ is relatively larger for the paired-atom component, and the system sits in the MI($\rho_a=0$, $\rho_p=1$) phase.
In the range $-1\textless\mu/U\textless-0.8$, the system is still at the MI($\rho_a=0$, $\rho_p=1$) phase.
The density $\rho_a\ne0$ is due to the finite size effects, which are eliminated by finite size scaling.
The scaling of $\rho_a$ with sizes $L=20,24, 48$ at $\mu/U=-0.9$ is done and
$\rho_a$ vanishes at the thermodynamic limit.
For the sake of simplicity,  the scaling is not shown here.

Similarly, in the range $-0.3\textless\mu/U\textless-0.1$, the system is in the MI ($\rho_a=1$, $\rho_p=1$) phase.
In the range $-0.8\textless\mu/U\textless-0.3$, the system is in the ASF phase as $C_a\ne0$ obviously.

In the plots of Fig.~\ref{N1J04}, we only show the results with size $L=48$, and
one  may therefore worry  about the size effect causing
the deviation of the phase transition point $\mu_c/U$ from the MI($\rho_a=0$, $\rho_p=1$) phase to the ASF phase. The finite size scaling of
the transition points $\mu_c/U$ shows that the deviation is invisible, as shown in Fig.~\ref{N1J04}(b).
% TCS: 'finite size scaling' is the subject, therefore singular.

The finite size effects of $C_a\ne0$ in both ranges  $-1\textless\mu/U\textless-0.8$ and  $-0.3\textless\mu/U\textless-0.1$
can be confirmed by the behaviors of $C_{a}(r)$.
In Fig.~\ref{N1J04}(c), the exponentially decaying function $C_{a}(r)$ satisfying $C_{a}(r)=a_{0}e^{-0.65(1)r}$ at $\mu/U=-0.9$, shows that there is no superfluid order and the system sits in the MI($\rho_a=0$, $\rho_p=1$) phase.\cite{decay1,decay2}
While at $\mu/U=-0.5$,  the algebraically decaying function $C_{a}(r)$, namely $C_{a}(r)=a_{0}r^{-0.26(1)}$, tells us that the system is of a strong superfluid order and sits in the ASF phase.

In the whole range$-2\textless\mu/U\textless1$,
$\rho_p$ maintains itself at unity. The reason why is
because $\mu_p$  is independent of $\mu$ and is a constant,
which results in that the paired-atom component sits in the MI($\rho_p=1$) phase.

The system includes the single atom and the pair atoms coupling term $p_{i}a_i^\dagger p_{i+1}^\dagger a_{i+1}$ in Eq.~(\ref{Hap}).
Compared to the usual hardcore BH model,\cite{hardcore} the phase diagram is
asymmetric about $\mu$.

\subsection{Softcore  off-site trimer model}
\subsubsection{The global phase diagram}
\vskip 0.15cm
\begin{figure}[hbt]
\begin{center}
\includegraphics[width=0.9\columnwidth]{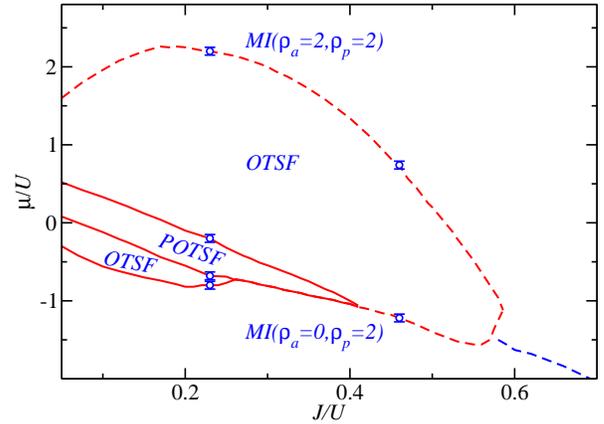}
\caption{(color online) The softcore phase diagram $(J/U-\mu/U)$ with the maximum occupation number $n_{max}^{a}=n_{max}^{p}=2$. The lines are schematic phase boundaries.  The dashed lines denote the first order phase transition, and the solid lines mark the continuous phase transition.}
\label{n2md}
\end{center}
\end{figure}
Since the OTSF phase doesn't emerge in the hardcore model, and Ref.~\cite{central} found the OTSF phase in the softcore model, we will study the softcore BH model in this section, and expect to find the stable OTSF phase.

Fig.~\ref{n2md} shows the global phase diagram of the softcore BH model. When $\mu/U$ is very large or negative, the system sits in the MI($\rho_a=2$, $\rho_p=2$) and MI($\rho_a=0$, $\rho_p=2$) phases, respectively.
Furthermore, in the regime where $J/U$ is relatively strong, the system is in the two phases.
Fortunately, in contradistinction to the hardcore model, when the ratio $J/U$ becomes weaker, the expected OTSF phase appears and distributes into a large region.
More interestingly, apart from the pure OTSF phase, a POTSF phase distributes between the two OTSF phases.
In this parameter regime, the single atom tunneling and the off-site tunneling
can exist simultaneously.
In other words, a part of the OTSF phase is broken by the on-site repulsions, and becomes a normal SF phase, i.e., a POTSF phase. It is similar to the partially paired (PP) phase in the anyon Hubbard model.\cite{pp, ppphase}

Furthermore, the phase transitions between the OTSF phase and other phases are also interesting.
The phase transition from the MI phase to the OTSF phase can be continuous or first order.
When $J/U\textless0.41$, it is continuous, and will turn into first order transition when $J/U\textgreater0.41$.
Moreover, the phase transition from the OTSF phase to the MI($\rho_a=2$, $\rho_p=2$) phase is the first order.
Furthermore, it should be noted that the phase transition between the POTSF phase and the pure OTSF phase is still continuous.
This phase transition is similar to the transition that from the PP phase to the SF or PSF phase in Ref.~\cite{pp}.
All these phase transitions were not discussed in Ref.~\cite{central}.

\subsubsection{Detailed analysis}
\begin{figure}[hbt]
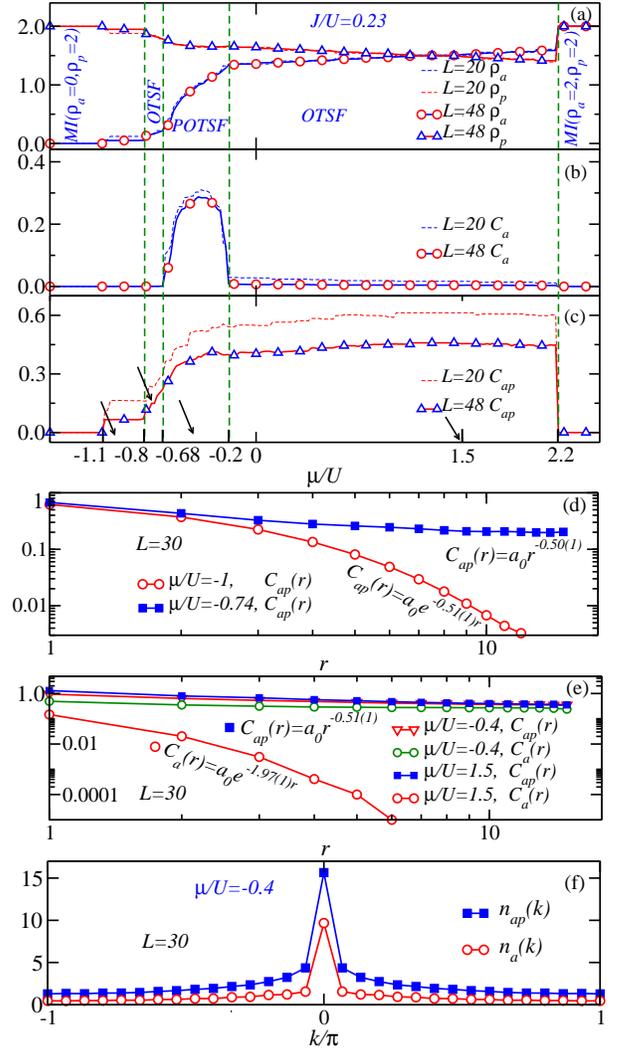
%\vskip -0.45cm \hskip -0.2cm
\begin{center}
\includegraphics[width=0.9\columnwidth]{fig6abc.eps}\\\hskip -0.1cm
\includegraphics[width=0.905\columnwidth]{fig6d.eps}\\
\includegraphics[width=0.905\columnwidth]{fig6e.eps}\\ \hskip -0.15cm
\hskip 0.23cm
\includegraphics[width=0.905\columnwidth]{fig6f.eps}
\caption{(color online) The values of the (a) densities $\rho_a$, $\rho_p$, (b) average correlation functions $C_a$, and (c) $C_{ap}$ vary with $\mu/U$ at $J/U=0.23$ with sizes $L=20$ and $48$.
(d) $C_{ap}(r)$ at $\mu/U=-1$ and $-0.74$, (e) $C_{ap}(r)$ and $C_{a}(r)$ at $\mu/U=-0.4$ and $1.5$, (f) $n_{a}(k)$ and $n_{ap}(k)$ at $\mu/U=-0.4$ with sizes $L=30$. }
\label{N2J023}
\end{center}
\end{figure}

\begin{figure}[htb]
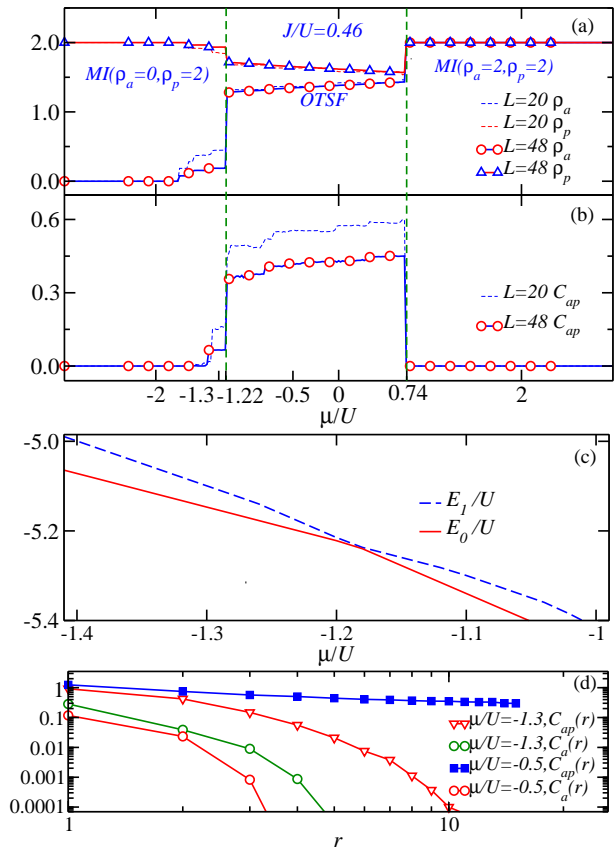

\begin{center}
\includegraphics[width=0.9\columnwidth]{fig7ab.eps}\\ \hskip -0.15cm
\includegraphics[width=0.905\columnwidth]{fig7c.eps}\\ \hskip -0.35cm
\includegraphics[width=0.925\columnwidth]{fig7d.eps}
\caption{(color online) The values of the (a) densities $\rho_a$, $\rho_p$, (b) average correlation functions $C_{ap}$ vary with $\mu/U$ at $J/U=0.46$ with sizes $L=20$ and $48$.
(c) $E_1/U$ and $E_0/U$ mean dimensionless excited and ground states energies, respectively, from the MI($\rho_a=0$, $\rho_p=2$) phase to the OTSF phase by exact diagonalization with four lattice sites. (d) $C_{ap}(r)$ and $C_{a}(r)$ at $\mu/U=-1.3$ and $-0.5$ with sizes $L=30$ and $48$, respectively. }
\label{N2J046}
\end{center}
\end{figure}

In order to analyze these phase transitions and give detailed descriptions of each physical quantity, we select $J/U=0.23$ and $0.46$ as examples, as shown
in Figs.~\ref{N2J023} and ~\ref{N2J046}, respectively.
%\newpage

In Fig.~\ref{N2J023}(a), when $\mu/U\textless-0.8$, the system sits in the MI($\rho_a=0$, $\rho_p=2$) phase.
There is no single atom on the lattice, and every site is occupied by two pairs of atoms, i.e., four atoms.
The reason why is because this scenario is similar to that of the hardcore model where $\mu_p$ is relatively large.
In the range $-1.1\textless\mu/U\textless-0.8$, there is a platform of $C_{ap}$ in Fig.~\ref{N2J023}(c), $C_{ap}\ne0$ with  two finite sizes $L=20$ and $48$.
However, It doesn't mean that the system sits in the OTSF phase, due to $C_{ap}(r)$ obeying an exponentially decaying curve rather than an algebraically decaying
curve in these parameter regimes, as shown at $\mu/U=-1$ in Fig.~\ref{N2J023}(d).

Continuously increasing $\mu/U$, in the narrow region $-0.8\textless\mu/U\textless-0.68$, an OTSF phase emerges, as $C_{ap}\ne0$ obviously, confirmed by the algebraically decaying $C_{ap}(r)$ at $\mu/U=-0.74$ in Fig.~\ref{N2J023}(d). It should be noted that $C_{a}(r)$ is not shown here due to the fact that $C_{a}=0$.
Moreover, the continuous variation of $C_{ap}$ implies that the phase transition from the MI($\rho_a=0$,$\rho_p=2$) phase to the OTSF phase is also continuous.
Similarly, the OTSF phase also emerges in the region $-0.2\textless\mu/U\textless2.2$. In Fig.~\ref{N2J023}(e), at $\mu/U=1.5$,  the
algebraically decaying $C_{ap}(r)$ and  exponentially decaying  $C_{a}(r)$ confirm that the system sits in a pure OTSF phase. Moreover, the quantity $C_{ap}$ jumps at $\mu/U=2.2$ in Fig.~\ref{N2J023}(c), with the implication that the phase transition between the OTSF phase and the MI($\rho_a=2$, $\rho_p=2$) phase is first order.\cite{foqpt1}

Apart from the pure OTSF phase, a POTSF phase also emerges. In the range $-0.68\textless\mu/U\textless-0.2$,
in Figs.~\ref{N2J023}(b) and (c),  both $C_{a}$ and $C_{ap}$ are obviously nonzero, and both $C_{ap}(r)$ and $C_{a}(r)$
obey algebraically decaying functions, as shown in Fig.~\ref{N2J023}(e), at $\mu/U=-0.4$. The phase transition from the POTSF phase to the OTSF phase
is continuous.

For cold atom experiments, the momentum distribution
can be used to detect the novel POTSF phase.
In Fig.~\ref{N2J023}(f), by a Fourier transformation of both $C_{ap}(r)$ and $C_{a}(r)$ according to Eq.~(\ref{momentum}), we can get the momentum distributions $n_{a}(k)$ and $n_{ap}(k)$, both of which have peaks at $k=0$, similar to the outcome of Ref.~\cite{central}.
According to the characteristics of the momentum distributions in Table 2, the  two peaks indicate that the system is indeed in the POTSF phase rather than any other phase.

Apart from $J/U=0.23$, the details concerning $J/U=0.46$ are discussed and $\rho_a$, $\rho_p$ and $C_{ap}$ are shown by scanning
$\mu/U$ in Figs.~\ref{N2J046}(a) and (b). Clearly, the phase transitions from the OTSF phase to both the MI($\rho_a=0$, $\rho_p=2$) or MI($\rho_a=2$, $\rho_p=2$) phases, satisfy the discontinuous change of a order parameter when a first-order phase transition takes place.\cite{foqpt1}

To understand the first-order phase transition more intuitively, the excited and ground energies are also calculated by exact diagonalization with four lattice sites, as shown in Fig.~\ref{N2J046}(c), where an intersection emerges around $\mu/U=-1.22$, which is consistent with the phenomenon of coexistence  between two phases at the critical phase transition point as the first-order phase transition\cite{foqpt2} occurs from the MI($\rho_a=0$, $\rho_p=2$) phase to the OTSF phase. A small deviation of this phase transition point is caused by the finite size effect.

In Fig.~\ref{N2J046}(b), near $\mu/U=-1.3$, $C_{ap}\ne0$ for the two finite sizes $L=20$ and $48$ as $C_{ap}$ is a platform in the range
$-1.1\textless\mu/U\textless-0.8$ in Fig.~\ref{N2J023}(c), yet we are still certain the system sits in the MI($\rho_a=0$, $\rho_p=2$) phase, based
on the fact that $C_{ap}(r)$ and $C_{a}(r)$ obey exponentially decaying functions in this region, as shown at $\mu/U=-1.3$ in Fig.~\ref{N2J046}(d). Furthermore, when $\mu/U=-0.5$, the algebraically decaying $C_{ap}(r)$ and exponentially decaying $C_{a}(r)$ confirm that the system sits in a pure OTSF phase. The reason why $C_{a}$ is not shown in Figs.~\ref{N2J046} is that $C_{a}(r)$ obeys exponential decay.

\section{Discussion and conclusion}
\label{sec:con}
By using the DMRG method, the BH model, with an effective off-site three-body tunneling and the attractive nearest-neighbor interaction, was studied
on a one-dimensional optical lattice.

The OTSF phase and POTSF phases of the softcore BH model, is stable in the thermodynamic limit.
The phase transition between the OTSF phase and the MI phase can be first order or continuous, depending on the parameters $J/U$ and $\mu/U$.

Our results numerically verify that the OTSF phase derived in the momentum space from Ref.~\cite{central} is stable in the thermodynamic limit. By using the DMRG method, we calculated the global phase diagrams and the
interesting phase transitions. This is an important extension of the mean field work in Ref.~\cite{central}.
The new results in this work concerning the OTSF and POTSF phases are helpful in realizing and understanding the novel  phases by cold atom experiments.
\addcontentsline{toc}{chapter}{Acknowledgment}
\section*{Acknowledgment}
We thank Zhao Ji-Ze, Ding Cheng-Xiang, Duan Cheng-Bo and Tang Gui-Xin for their invaluable discussions.
T.C. Scott was supported in China by the project GDW201400042 for the ``high end foreign experts project''.
W.Z. Zhang was supported supported by the National Natural Science Foundation of China (Grant Nos.~11305113).

\appendix
\end{CJK*}
\end{document}